\documentclass[final,3p,times,twocolumn]{elsarticle}

\usepackage{amssymb}
\usepackage{amsmath,amssymb,exscale}
\usepackage{graphicx}
\usepackage{epsfig}
\usepackage{multicol}
\usepackage{color}
\usepackage{xcolor}
\usepackage{hyperref}
\usepackage{mathrsfs}
\usepackage{hyperref}
\hypersetup{colorlinks,bookmarksopen,bookmarksnumbered,citecolor=rossos,
linkcolor=black,pdfstartview=FitH,urlcolor=blus}
\usepackage{amsfonts}
\usepackage{slashed}
\usepackage{blindtext}
 \usepackage{fancyhdr}
\usepackage{hyperref}
\usepackage{mathtools}
\biboptions{numbers,sort&compress}

\allowdisplaybreaks

\definecolor{blus}{cmyk}{1,1,0,0.}
\definecolor{verdes}{cmyk}{0.99,0,0.99,0.02}
\definecolor{rossos}{cmyk}{0,1,1,0.55}
\definecolor{greeny}{cmyk}{0.99,0,0.59,0.98}
\definecolor{redy}{cmyk}{0,1,1,0.40}

\newcommand{\bp}{M_P}

\def\be{\begin{equation}}
\def\ee{\end{equation}}
\def\bea{\begin{eqnarray}}
\def\eea{\end{eqnarray}}

\def\ba{\begin{array} }
\def\ea{\end{array}}
\def\bac{\begin{array} {c}}
\def\bacc{\begin{array} {cc}}
\def\baccc{\begin{array} {ccc}}

\def\psl{\hbox{\hbox{${p}$}}\kern-1.9mm{\hbox{${/}$}}}
\def\dsl{\hbox{\hbox{${\partial}$}}\kern-1.7mm{\hbox{${/}$}}}
\def\Dsl{\hbox{\hbox{${D}$}}\kern-2.1mm{\hbox{${/}$}}}

\definecolor{red}{rgb}{1,0,0}

\def\hhref#1{\href{http://arxiv.org/abs/#1}{arXiv:#1}}  
\hypersetup{colorlinks=true, linkcolor=red}

\journal{Physical Review D}

\begin{document}

\begin{frontmatter}

\title{\vspace{-2cm}\huge {\color{redy}Inflating and Reheating the Universe \\
with an Independent Affine Connection
}}

\author{\vspace{1cm}{\large {\bf Alberto Salvio}}}

\address{\normalsize \vspace{0.2cm}Physics Department, University of Rome and INFN Tor Vergata, Italy \\

\vspace{0.3cm} 
 \vspace{-1cm}
 }

\begin{abstract}
It is shown that a component of the dynamical affine connection, which is independent of the metric, can drive inflation in agreement with observations. This provides a geometrical origin for the inflaton. It is also found that the decays of this field, which has spin 0 and odd parity, into Higgs bosons can reheat the universe up to a sufficiently high temperature.
\end{abstract}


\end{frontmatter}
 
 \begingroup
\hypersetup{linkcolor=blus}
\tableofcontents
\endgroup


\section{Introduction}\label{introduction}
  
  \vspace{-0.2cm}
 
 Einstein's general relativity (GR) explains gravity in geometrical terms, the distances are measured through the metric and the gravitational force is determined by the (affine) connection, which is the essential building block of covariant derivatives. This beautiful construction accounts for all gravitational observations performed so far, including  today's nearly-exponential accelerated expansion of the universe if the cosmological constant is present. 
 
 It is generically accepted that another, but much more rapid, nearly-exponential expansion occurred during the early stages of the universe (inflation). This can be driven by a spin-0 field, the inflaton, with an appropriate potential, which guarantees  that such an expansion not only occurred, but also  eventually came to an end. Indeed, a reheating  must take place after inflation in order to generate all particles we observe.
 
 From the purely geometrical point of view the metric and the connection, unlike in GR, can be completely independent objects and, moreover, can contain extra degrees of freedom besides the spin-2 graviton. This generalized scenario is known as metric-affine gravity (see~\cite{Baldazzi:2021kaf} for a recent discussion and further references).
 
The goal of this paper is to discover whether the role of the inflaton can be played by an extra dynamical component of the connection. The main motivation behind this goal is to provide a geometrical origin for the inflaton too, linking it to one of the essential geometrical objects, the connection. In order to achieve this goal not only an inflaton with an appropriate potential should be identified among the components of the connection, but also the current constraints given by cosmic microwave background (CMB) observations should be satisfied. These are provided by Planck~\cite{Ade:2015lrj} and, more recently, the BICEP and Keck collaborations~\cite{BICEP:2021xfz}. Moreover, as discussed, the universe must be appropriately reheated after inflation. This requires an efficient production of known particles, such as electrons, quarks and Higgs bosons.

In the following sections we show that all this is possible and we work out the predictions in a simple, yet well-motivated, model.  
 
  \vspace{-0.1cm}
 
\section{The key idea and inflation}\label{Inflation}

  \vspace{-0.2cm}
  
When the connection ${\cal A}_{\mu~\sigma}^{~\,\rho}$ and the metric, $g_{\mu\nu}$, are independent there are two rather than one invariant that are linear in the curvature
\be {\cal R}_{\mu\nu~~\sigma}^{~~~\rho} \equiv \partial_\mu{\cal A}_{\nu~\sigma}^{~\,\rho}+{\cal A}_{\mu~\lambda}^{~\,\rho}{\cal A}_{\nu~\sigma}^{~\,\lambda}- (\mu\leftrightarrow \nu).\ee
The first one is the usual Ricci-like scalar\footnote{Greek indices are lowered and raised by $g_{\mu\nu}$.} ${\cal R}\equiv {\cal R}_{\mu\nu}^{~~~\mu\nu}$ and the second one is the parity-odd Holst invariant ${\cal R'}\equiv \epsilon^{\mu\nu\rho\sigma}{\cal R}_{\mu\nu\rho\sigma}/\sqrt{-g}$~\cite{Hojman:1980kv,Nelson:1980ph,Holst:1995pc}, where $\epsilon^{\mu\nu\rho\sigma}$ is the totally antisymmetric Levi-Civita symbol with $\epsilon^{0123}=1$ and $g$ is the determinant of the metric. In the GR case, where ${\cal A}_{\mu~\sigma}^{~\,\rho}$ equals the Levi-Civita connection, ${\cal R}$ coincides with the Ricci scalar, $R$, but ${\cal R'}$ vanishes. For this reason in metric-affine gravity ${\cal R'}$ can be understood as a component of the connection. 

The key idea here is to identify the inflaton with ${\cal R'}$. To do so ${\cal R'}$ has to be a dynamical field, which is independent of the metric, and the simplest inflationary action that realizes this is
\be S_I= \int d^4x\sqrt{-g}\left( \alpha{\cal R}+\beta {\cal R'} + c {\cal R'}^2\right). \label{SI}\ee
Indeed, for $c=0$ one can easily show, by solving the connection equations, that $S_I$ is equivalent to the Einstein-Hilbert action for any $\beta$, having identified $\alpha=\bp^2/2$, where $\bp$ is the reduced Planck mass. While for $c\neq0$, standard auxiliary field methods show that an extra spin-0 parity odd dynamical field $\zeta'$, which is introduced as an auxiliary field, is present and precisely equals  ${\cal R'}$ on shell~\cite{Hecht:1996np,BeltranJimenez:2019hrm,Pradisi:2022nmh}: we can equivalently write
\be S_I=  \int d^4x\sqrt{-g}\left[ \alpha{\cal R}+\left(\beta+2c\zeta' \right) {\cal R'} -c\zeta'^2\right],\ee 
which coincides with the expression in~(\ref{SI}) after using the $\zeta'$ equation. Because of its symmetry properties we call $\zeta'$ the pseudoscalaron. The $\beta {\cal R'}$ term, known as the Holst term, is also necessary to obtain a suitable inflaton potential, as we will see; the quantity $\bp^2/(4\beta)$ is called the  Barbero-Immirzi parameter~\cite{Immirzi:1996di,Immirzi:1996dr}. After using the ${\cal A}_{\mu~\sigma}^{~\,\rho}$ equation a non-canonical kinetic term of $\zeta'$ appears.
 It is possible to canonically normalize the pseudoscalaron by considering the field redefinition 
  \be  \zeta'(\omega) = \frac1{2c}\left(\frac{\bp^2 \tanh X(\omega)}{4\sqrt{1-\tanh^2X(\omega)}}-\beta\right),\label{zetapomega}\ee
 where
\be X(\omega)\equiv \sqrt{\frac{2}{3}}\frac{\omega}{\bp}+\tanh ^{-1}\left(\frac{4 \beta }{\sqrt{16 \beta ^2+\bp^4}}\right), \label{Xofomega}  \ee such that, after using the connection equations, $S_I$ becomes a standard scalar-tensor action~\cite{Pradisi:2022nmh} (we use the mostly plus convention for the metric)
   \be S_I = \int d^4x\sqrt{-g}\left[\frac{M_P^2}{2} R-\frac{(\partial \omega)^2}{2} -U(\zeta'(\omega)) \right],\label{SeqCan}\ee
   where
     $U(\zeta'(\omega)) = c\zeta'(\omega)^2$ and clearly $c>0$ for stability reasons. As clear from~(\ref {SeqCan}) this model does not contain any ghost (the Einstein-Hilbert term has the usual sign and the kinetic term of $\omega$ contributes positively to the kinetic energy). Furthermore, for $c>0$ the  mass of $\omega$ (defined as the mass of the fluctuations of this field around a Lorentz invariant solution) is positive, namely $\omega$ is not a tachyon. The potential $U(\zeta'(\omega))$ is symmetric in the exchange $\{\omega, \beta\} \to \{-\omega, -\beta\}$. Therefore, an arbitrary value of $\beta$ and its opposite are physically equivalent. 
   
   The $c{\cal R'}^2$ term, besides being the simplest one leading to an extra spin-0 field, is also motivated by scale invariance and Weyl invariance at high energies: by replacing $g_{\mu\nu}\to \Omega^2 g_{\mu\nu}$ that term  is invariant, not only when $\Omega$ is spacetime independent (scale invariance), but also when it is spacetime depend (Weyl invariance). This is because in metric-affine gravity the metric and the connection are independent and a rescaling of the metric does not imply any change in the connection. The extension of the present model to a fully scale invariant one is beyond the scope of the present work because is not mandatory to assess the viability of this scenario: mass scales can also be added by hand. We thus leave such an extension as an interesting outlook for future work. This may be realized  perhaps along the lines of~\cite{Shaposhnikov:2008xi} (where the vacuum expectation value of a scalar field generates the mass scales, in our case $\alpha$ and $\beta$) or~\cite{Salvio:2014soa} (where the mass scales are induced through a gravitational version of the Coleman-Weinberg mechanism~\cite{Coleman:1973jx}). It is also interesting to note that the same inflationary predictions generically emerge if one substitutes $c{\cal R'}^2$ with a general quadratic function of both ${\cal R}$ and ${\cal R'}$, which is still compatible with scale invariance. This is because such more general function leads to the same potential, as recently shown in~\cite{Pradisi:2022nmh}. The inflationary predictions that we will find are, therefore, quite robust.
   
   The slow-roll approximation can be used when 
\be \epsilon \equiv\frac{\bp^2}{2} \left(\frac{1}{U}\frac{dU}{d\omega}\right)^2\ll 1, \quad \eta \equiv \frac{\bp^2}{U} \frac{d^2U}{d\omega^2}\ll 1 \label{epsilon-def}\ee
and in this case the number of e-folds $N_e$ as a function of the field $\omega$ is given by 
  \be N_e(\omega) = N(\omega)  - N(\omega_{\rm end}),    \ee
  where  \be N(\omega)  = \frac1{\bp^2}  \int^\omega d\omega'  \,  U\left(\frac{dU}{d\omega'}\right)^{-1}  \ee 
and $\omega_{\rm end}$ satisfies $\epsilon(\omega_{\rm end}) = 1$ (see details below).
 The scalar spectral index $n_s$, the tensor-to-scalar ratio $r$ and the curvature power spectrum  $P_R$ (at horizon exit) are then given by
\be n_s = 1- 6\epsilon +2\eta, \quad r =16\epsilon, \quad P_R= \frac{U/ \epsilon}{24\pi^2 \bp^4} \label{ns-r-PR}.\ee
 One finds analytic expressions not only for $\epsilon$, $\eta$, $n_s$, $r$ and $P_R$, but also for the e-fold functions $N$ and $N_e$ (see~\ref{appendix}). Indeed, the equation $\epsilon(\omega_{\rm end}) = 1$ can be solved for real $\omega_{\rm end}$ whenever $192 \beta^2\geq4\bp^2$ and one finds two solutions, which we call $\omega_\pm$ and whose analytic expression is given in Appendix~\ref{appendix}.
 Note that, as always the case in slow-roll inflation, $\epsilon$, $\eta$, $n_s$ and $r$ are independent of the overall constant in the potential ($1/c$ in this case), while $P_R$ is proportional to it. So the observed value of $P_R$ (namely $(2.10 \pm 0.03) \times 10^{-9}$~\cite{Ade:2015lrj})  can always be obtained by choosing $c$ appropriately. 
 
\begin{figure}[t!]
\begin{center} 
\hspace{-0.37cm} 
\vspace{0.4cm}
\includegraphics[scale=0.33]{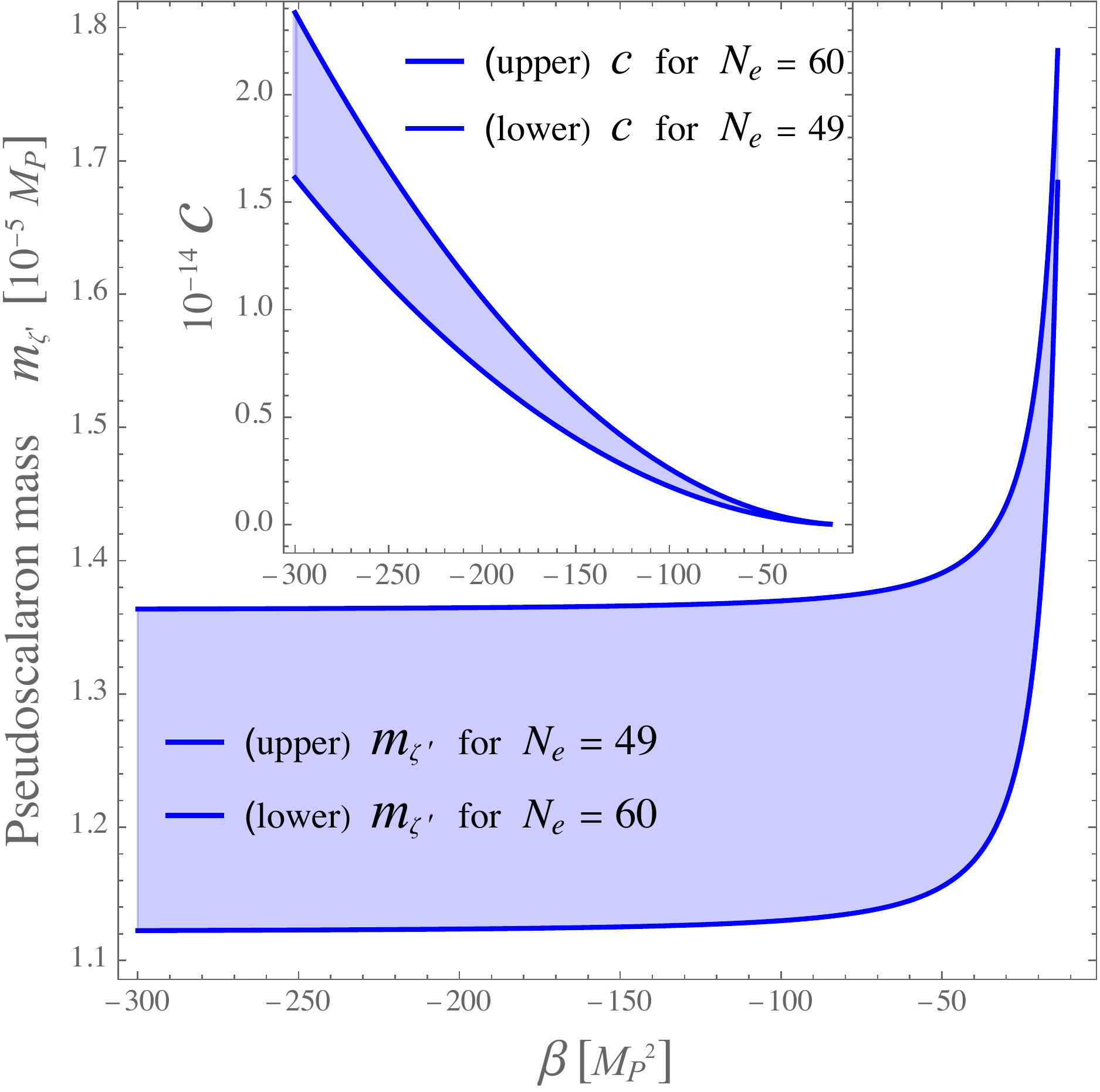}
\\
\hspace{-0.25cm}  \includegraphics[scale=0.33]{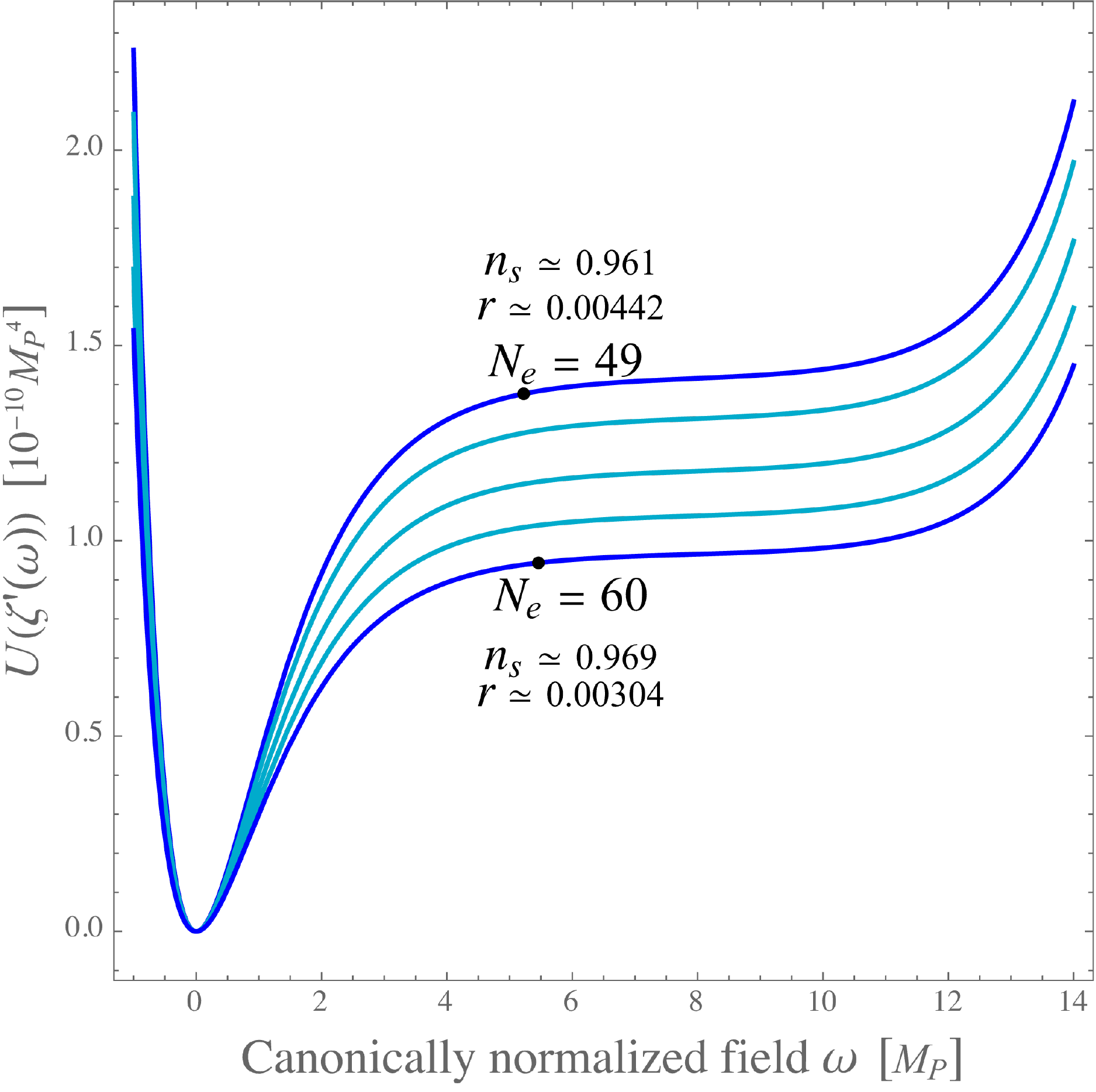}
 \end{center} 
   \caption{\em {\bf Upper plot:} the pseudoscalaron mass and the corresponding value of $c$ (in the inset) that gives $P_R=2.1 \times 10^{-9}$~\cite{Ade:2015lrj} $N_e$ e-folds before the end of inflation as a function of $\beta$. {\bf Down plot:} the corresponding pseudoscalaron potential for $\beta=-80\bp^2$. For all curves  $P_R=2.1 \times 10^{-9}$ by construction. Also, the black points correspond to the values of the inflaton for which $N_e=49$ (upper curve) and $N_e=60$ (lower curve) are realized; the corresponding predictions for $n_s$ and $r$ (in good agreement with the Planck, BICEP and Keck observations) are provided. }
\label{U-c}
\end{figure}

 We want $\omega_{\rm end}$ to be the value of $\omega$ at the end of inflation so, given the shape of $U(\zeta'(\omega))$, we take $\omega_{\rm end}$ such that $|\omega_{\rm end}|=\min(|\omega_+|,|\omega_-|)$.
 In Fig.~\ref{U-c}   we show the potential of $\omega$  (down plot) and its  mass $m_\omega=m_{\zeta'}$ (upper plot) by setting $c$ in a way that $P_R=2.1 \times 10^{-9}$ at $N_e$ e-folds before the end of inflation. In the $\omega$ potential there is a plateau, which increases for larger $|\beta|$ and disappears when $\beta=0$. This is the reason why the $\beta {\cal R'}$ term in $S_I$ is necessary. In the down plot of Fig.~\ref{U-c} $|\beta| = 80\bp^2$ is chosen and is enough to even have     60 e-folds.

\begin{figure}[t]
\begin{center} 
\hspace{-0.37cm}
  \includegraphics[scale=0.35]{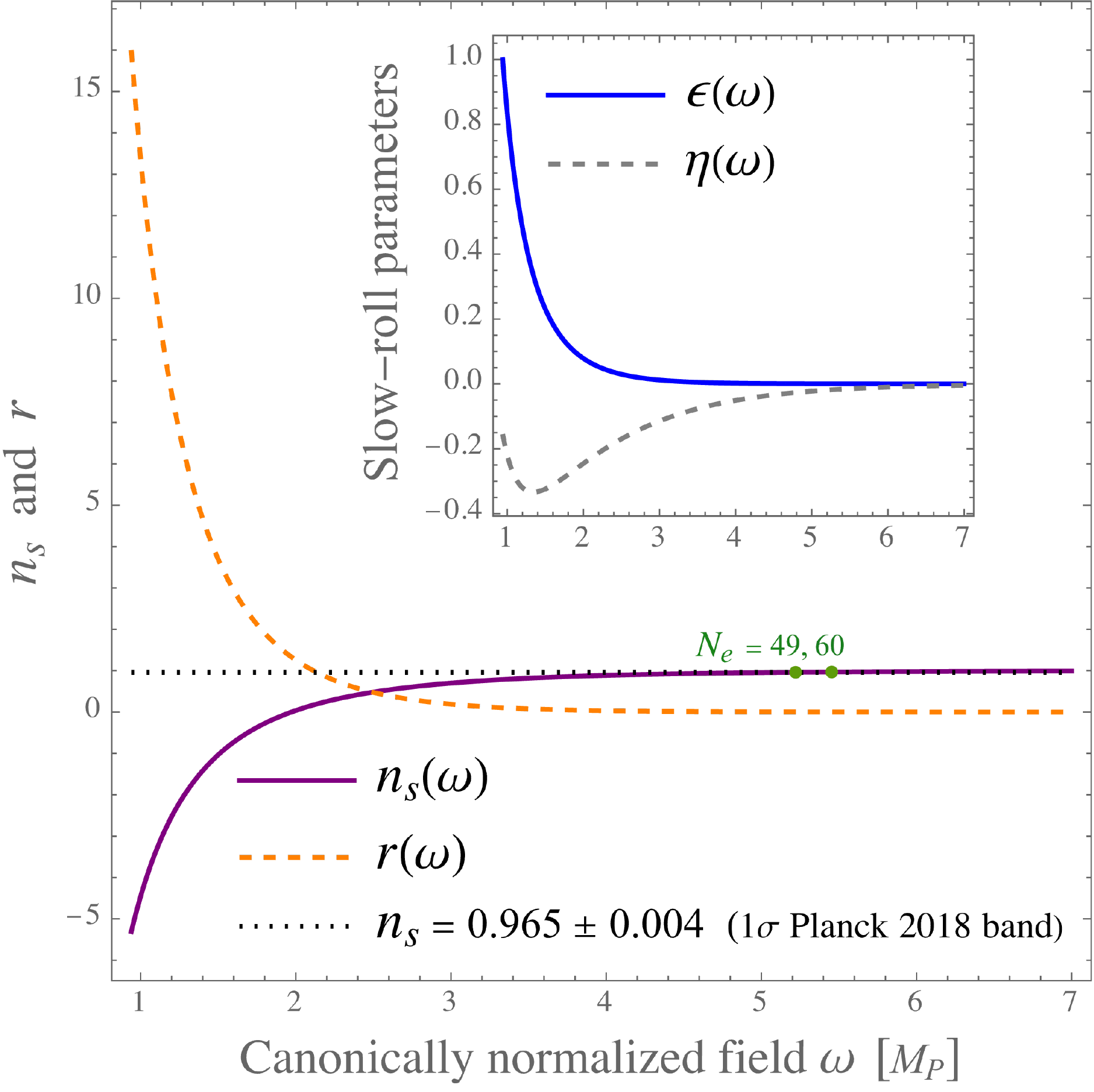} 
 \end{center} 
   \caption{\em  The scalar spectral index   and the tensor-to-scalar ratio  as functions of the canonically normalized pseudoscalaron. The pseudoscalaron values corresponding to $N_e =49, 60$ e-folds before the end of inflation are explicitly indicated: they correspond to the green points below the numbers $49$ and $60$, respectively. In the inset the slow-roll parameters are shown.  We have set $\beta = -300 \bp^2$.}
\label{inflationp}
\end{figure}

\begin{figure}[!t]
\begin{center}
\hspace{-0.37cm}
 \includegraphics[scale=0.345]{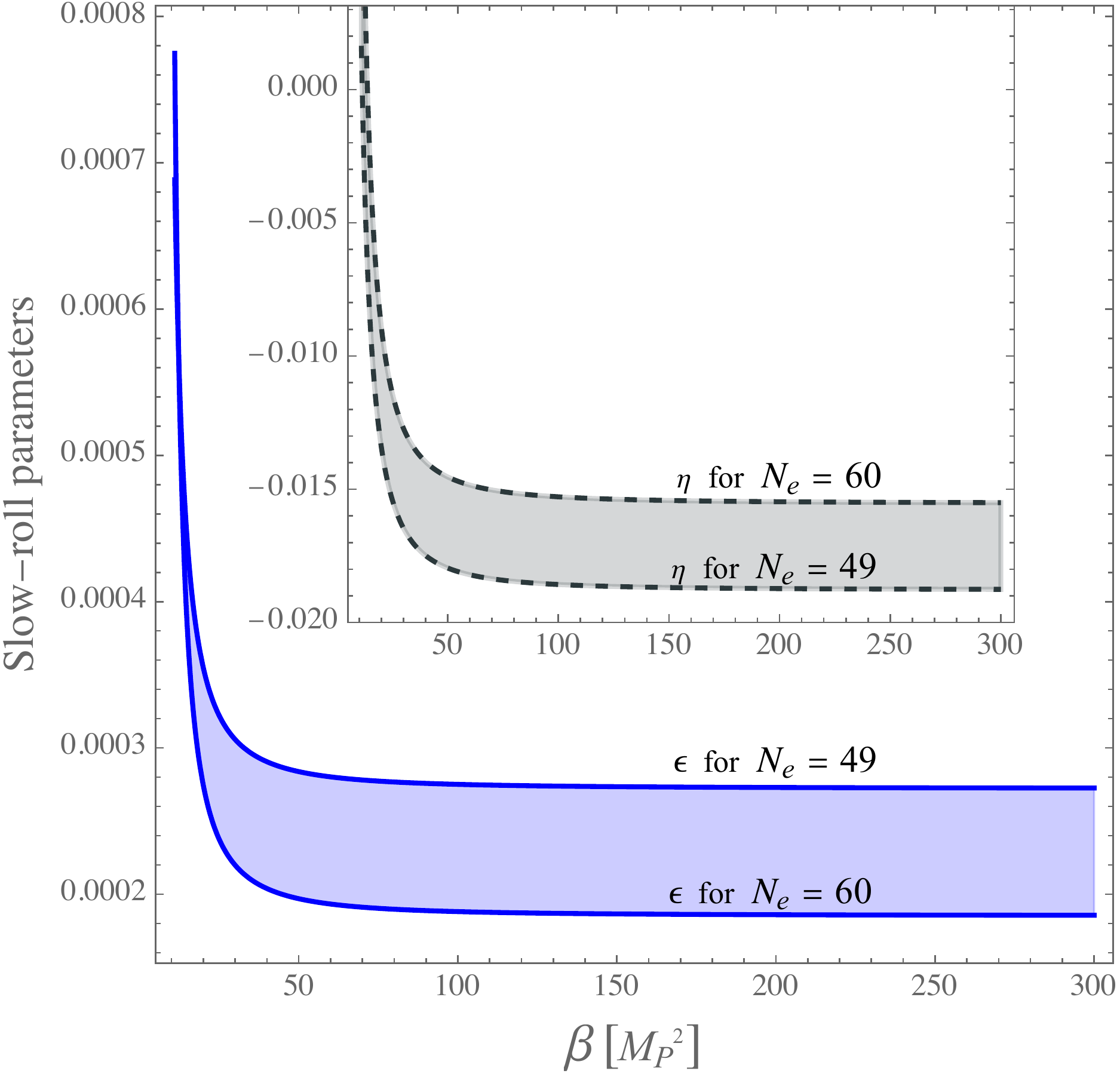} \\ 
 \vspace{0.16cm}
 \includegraphics[scale=0.345]{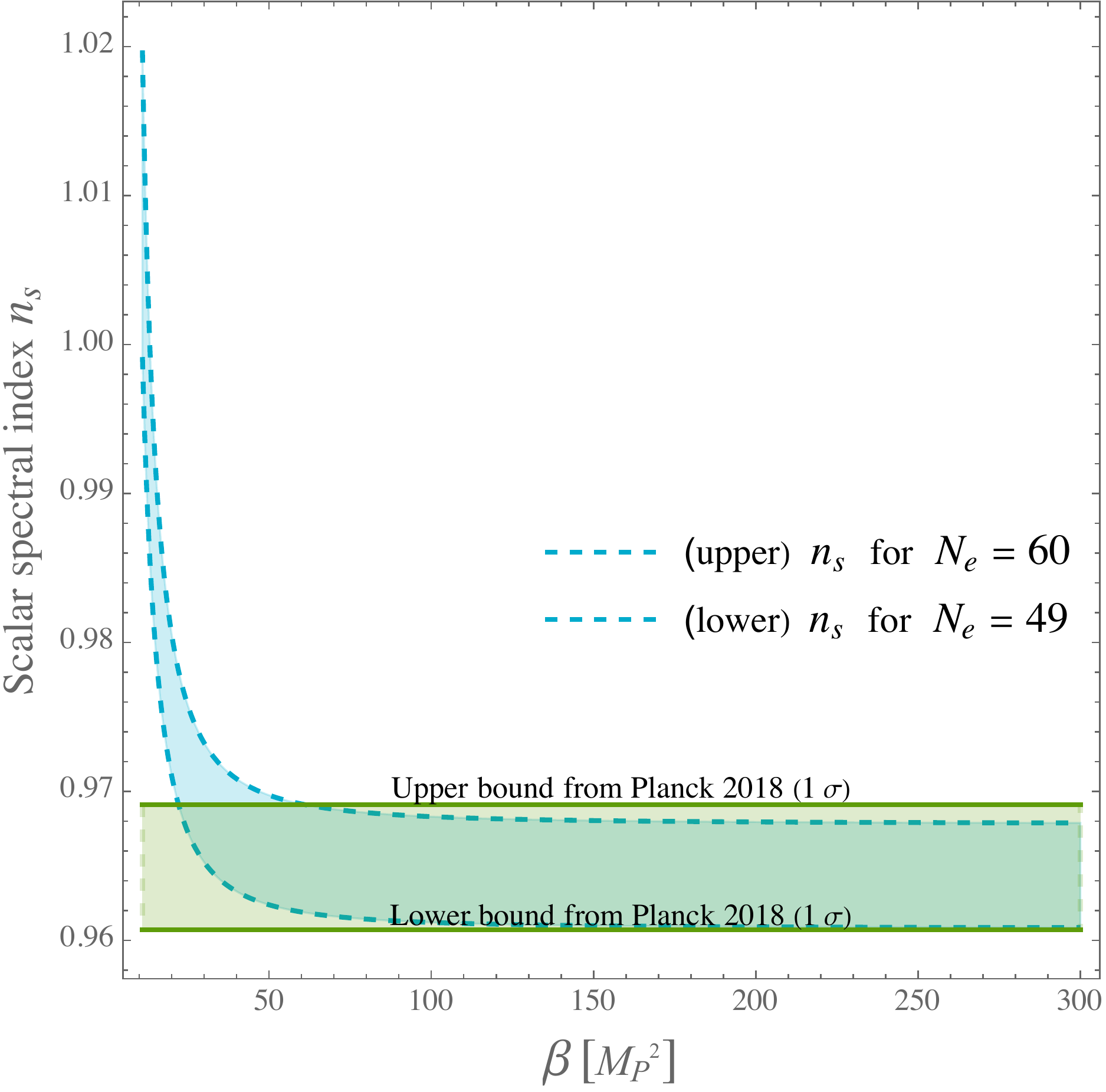}  \\
  \vspace{0.16cm}
  \includegraphics[scale=0.345]{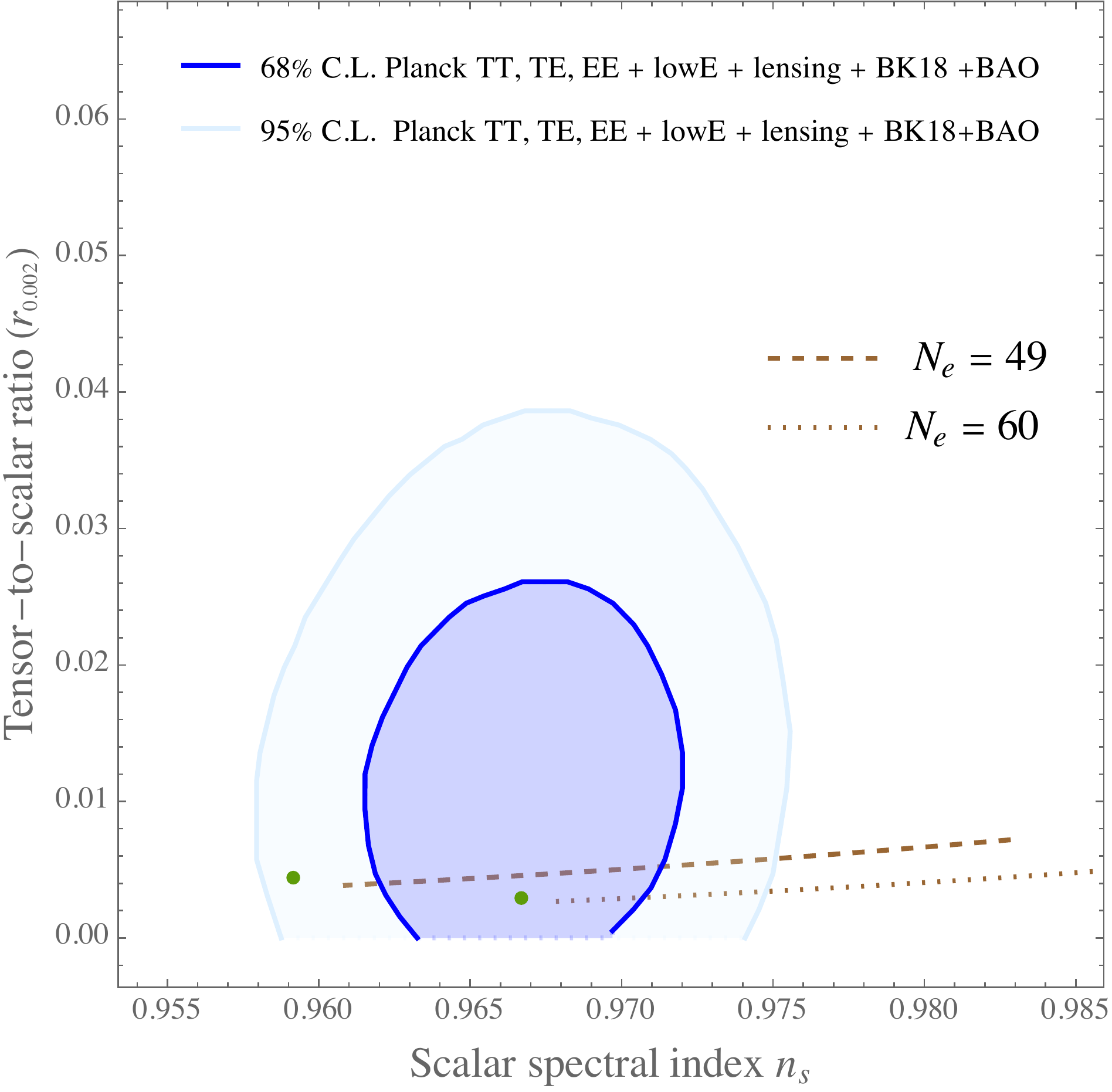}
 \end{center} 
   \caption{\em The slow-roll parameters,   the scalar spectral index   and the tensor-to-scalar ratio  as functions of $\beta$. The green dots in the bottom plot are the predictions of Starobinsky inflation.}
\label{inflation}
\end{figure}

In Figs.~\ref{inflationp} and~\ref{inflation} it is shown  that slow-roll inflation not only occurs, but is also  remarkably compatible with the most recent CMB observations provided by Planck  and BICEP/Keck (BK18 henceforth) for large $|\beta|$ (i.e.~small values of the Barbero-Immirzi parameter) and for an appropriate number of e-folds $N_e$~\cite{Ade:2015lrj}. Fig.~\ref{inflationp} shows that viable slow-roll inflation with an appropriate $N_e$ occurs for $\omega$ slightly above the Planck scale\footnote{However, the corresponding values of the energy density, $\sim U$, is well below the cutoff that is around the Planck scale (see the down plot of Fig.~\ref{U-c}).}; in that figure $\beta = -300 \bp^2$. In Fig.~\ref{inflation} we compare the observations and the theoretical predictions as functions of 
$\beta$ and show that viable slow-roll inflation with $N_e\simeq 49$ occurs already for $|\beta| \gtrsim 20\bp^2$ and with $N_e\simeq 60$ for $|\beta| \gtrsim 60\bp^2$. These values of the mass parameter $\sqrt{|\beta|}$ are above $\bp$, but not much larger than $\bp$: for $N_e\simeq 49$ and $N_e\simeq 60$ we have respectively $\sqrt{|\beta|} \gtrsim 4\bp$ and $\sqrt{|\beta|} \gtrsim 8\bp$. In that figure $r_{0.002}$ is the value of $r$ at the reference momentum scale $0.002~{\rm Mpc}^{-1}$, used by Planck  and BK18. In Figs.~\ref{inflation} we also report the predictions of Starobinsky inflation\footnote{In Starobinsky inflation the inflationary action $S_I$ also features a quadratic-in-curvature term, $S_I = \int d^4x \sqrt{-g}\left(\bp^2 R/2 + c R^2\right)$, but with connection equal to the Levi-Civita one. So it is interesting to compare the predictions of pseudoscalaron inflation with those of Starobinsky inflation.}~\cite{Starobinsky:1980te} for $n_s$ and $r$; the predictions of pseudoscalaron inflation approach (but do not quite reach) those of Starobinsky inflation for $|\beta|\to \infty$, while for a finite value of $\beta$ they differ significantly.

It is interesting to note that the predictions for $n_s$ and $r$ of pseudoscalaron inflation are within the reach of the future space mission LiteBIRD~\cite{LiteBIRD:2022cnt}, which will be, therefore, able to test this scenario. 

  \section{Reheating}\label{Reheating}
  \vspace{-0.2cm}
 Reheating the universe after inflation is mandatory for the viability of any model and, to achieve this, 
couplings between the inflaton  
and the Standard Model (SM) particles are needed. If $\omega$ decays into some SM particles with width $\Gamma_\omega$ the reheating temperature $T_{\rm RH}$ is at least
\be T_{\rm RH}\gtrsim \min\left( \left(\frac{45 \Gamma^2_{\omega}\bp^2}{4\pi^3 g_*}\right)^{1/4},\left(\frac{30 \rho_{\rm vac}}{\pi^2 g_*}\right)^{1/4}\right),\ee
where $g_*$ is the effective number of relativistic species in thermal equilibrium at temperature $T_{\rm RH}$ and $\rho_{\rm vac}$ is the vacuum energy density due to $\omega$ (note that $\rho_{\rm vac}$ represents the full energy budget of the system). This is the standard perturbative contribution to reheating; it is important to keep in mind  that there may also be non-perturbative contributions to the particle production after inflation~\cite{Dolgov:1989us,Traschen:1990sw,Kofman:1994rk,Kofman:1997yn}.   
However, we leave their detailed calculation for future work because
they are not crucial to assess the viability of the present scenario. This is because, as we will see, the value of $T_{\rm RH}$ estimated through the standard perturbative approach can be large enough.

Let us first consider a fermion $f$ represented by a Dirac spinor $\Psi$ minimally coupled to gravity and with mass $m_f$, i.e. with action
\be S_f = \int \sqrt{-g} \, \frac12 \overline \Psi (i \slashed{\cal D} -m_f)\Psi +{\rm h.c.}\, ,\ee 
where $\slashed{\cal D} \Psi \equiv \gamma^a e_a^\mu {\cal D}_\mu \Psi$  (the $e_a^\mu$ satisfy~\cite{etaab} $e_a^\mu e_b^\nu g_{\mu\nu}=\eta_{ab}$), $\overline \Psi \equiv \Psi^\dagger \gamma^0$, the Dirac gamma matrices $\gamma^a$ satisfy $\{\gamma^a,\gamma^b\} = -2\eta^{ab}$, 
${\cal D}_\mu \Psi = \partial_\mu \Psi +{\cal A}^{ab}_\mu[\gamma_a,\gamma_b]\Psi/8$ and ${\cal A}_{\mu~b}^{~\,a}  = e^a_\nu{\cal A}_{\mu~\lambda}^{~\,\nu} e^\lambda_b- e^\lambda_b\partial_\mu e^a_\lambda$. By using the connection equations with the formalism of~\cite{Pradisi:2022nmh}, one finds the following effective pseudoscalaron-fermion-fermion interaction 
 \be \mathscr{L}_{\omega ff} =   \frac{c_{\omega ff}}{\bp} \, \partial_\mu\omega\,\overline\Psi \gamma_5 \gamma^\mu\Psi,\ee
 where 
 \be c_{\omega ff} = \left[\frac{3\bp}{1+16 B^2}\frac{dB}{d\omega} \right]_{\omega=0} =\sqrt{\frac{3\bp^4}{8(\bp^4+16 \beta ^2)}},\ee
 $\gamma_5=i\gamma_0\gamma_1\gamma_2\gamma_3$ and $B =(\beta+2c\zeta'(\omega))/\bp^2$.    
 This effective interaction leads to the decay $\omega\to ff$ with width
 \be \Gamma_{\omega\to ff} = |c_{\omega ff}|^2\frac{m_{\omega} m_f^2}{2\pi \bp^2} \sqrt{1-\frac{4m^2_f}{m_\omega^2}}. \ee
 This channel can efficiently reheat the universe up to a temperature above the electroweak scale if $m_f$ is very large compared to that  scale. Such a fermion is not present in the SM. It is possible to engineer a model where there is a very heavy fermion with sizable couplings to SM particles such that this channel is sufficient.  For example, this is the case in the well-motivated model~\cite{Kim:1979if,Shifman:1979if}, which was proposed to solve the strong CP problem.

 However, in order to keep our analysis as model independent as possible, we consider another channel: the decay of $\omega$ into two identical real scalar particles, e.g. two Higgs bosons. This channel can be active when there is a non-minimal coupling between the real (canonically normalized) scalar field $\phi$ in question and ${\cal R}$  in the action: 
 \be S_{\rm nm}=\int \sqrt{-g}\frac{\xi \phi^2}{2} {\cal R}. \label{Snm} 
  \ee 
 This term is known to be generated by quantum corrections and, therefore, it is more natural to include it.
 If one solves the connection equation  in the presence of~(\ref{Snm}) (using the results in~\cite{Pradisi:2022nmh}) one finds the following effective pseudoscalaron-scalar-scalar interaction
  \be \mathscr{L}_{\omega \phi\phi} =   \frac{c_{\omega \phi\phi}}{\bp} \, \partial_\mu\omega\,\phi \partial^\mu\phi,\ee
  where 
 \be c_{\omega \phi\phi} = \left[\frac{48\xi \bp B}{1+16 B^2} \,\frac{dB}{d\omega}\right]_{\omega=0}=\frac{4 \sqrt{6} \beta  \xi }{\sqrt{\bp^4+16 \beta ^2}}.\ee  
 This effective parity-violating operator only arises in the presence of the  Holst term  because $c_{\omega \phi\phi}\to 0$ as $\beta\to 0$.
 The effective interaction $\mathscr{L}_{\omega \phi\phi}$ leads to the decay $\omega\to \phi\phi$ with width
 \be \Gamma_{\omega\to \phi\phi} = |c_{\omega \phi\phi}|^2\frac{m_{\omega}^3}{16\pi \bp^2} \sqrt{1-\frac{4m^2_\phi}{m_\omega^2}}, \label{HiggsChannel}\ee
 where $m_\phi$ is the mass of $\phi$. The produced Higgs particles subsequently decay into other SM particles, such as leptons and quarks. The channel $\omega\to \phi\phi$ can efficiently and naturally reheat the universe up to a temperature much above the electroweak scale, even if one identifies $\phi$ with the SM Higgs, so {\it per se} it does not  require any beyond-the-SM physics. For example, taking $m_\phi\ll m_\omega$, $g_*\sim 10^2$ and $\beta \gtrsim \bp^2$  one finds $T_{\rm RH}\gtrsim 10^9 |\xi|$~GeV.  This reheating temperature is compatible with all numbers of e-folds considered in Sec.~\ref{Inflation} for natural values of $|\xi|$ of order 1 or smaller.  Since $c_{\omega \phi\phi}\to 0$ as $\beta\to 0$ this reheating channel occurs thanks to the presence of an independent connection: the Holst term would be absent if the full connection were exactly the Levi-Civita one.
  \vspace{-0.1cm}
   
\section{Conclusions}\label{Conclusions}
  \vspace{-0.2cm}
It has been found that a pseudoscalar component of a dynamical connection, which is independent of the metric, can drive inflation in agreement with current data. This pseudoscalaron is identified with the parity odd Holst invariant and inflationary predictions in excellent agreement with data have been found for small values of the Barbero-Immirzi parameter, where the inflaton potential develops a plateau. The predictions approach, but do not quite reach, those of Starobinsky inflation as the Barbero-Immirzi parameter goes to zero; for finite values, on the other hand, the predictions significantly differ. Pseudoscalaron inflation can be tested by future CMB observations, such as those of LiteBIRD.

Moreover, the decays of the pseudoscalaron into Higgs particles (which occur thanks to the presence of an independent connection) can efficiently reheat the universe after inflation up to a high enough temperature. This temperature could be further increased by other channels, such as decays into very massive fermions, which we have computed too.

As an outlook example, it would be interesting to calculate the non-perturbative particle production after inflation (preheating). Moreover, it would also be interesting to engineer a fully scale invariant version of this model. Indeed, a crucial ingredient of the present construction is a quadratic-in-curvature term, $c{\cal R'}^2$, which is compatible with scale (and even Weyl) invariance.

\vspace{-0.2cm}

\section*{Acknowledgments} 
\vspace{-0.1cm} 

\noindent   I thank G.~Pradisi 
  for useful discussions. This work has been partially supported by the grant DyConn from the University of Rome Tor Vergata.

\vspace{0.cm}

\appendix

\section{\\ Analytic expressions for inflationary quantities}\label{appendix}

The analytic expressions for $\epsilon$, $\eta$, $n_s$, $r$,$P_R$ and $N$ are
{\allowdisplaybreaks\bea\epsilon(\omega)&=& \frac{4 \bp^4 \cosh^2X(\omega)}{3 \left(\bp^2 \sinh X(\omega)-4 \beta \right)^2}, \nonumber \\ 
\eta(\omega) &=& \frac{4 \bp^2 \left(\bp^2 \cosh \left(2 X(\omega)\right)-4 \beta  \sinh X(\omega)\right)}{3 \left(\bp^2 \sinh X(\omega)-4 \beta \right)^2},\nonumber
\\
N(\omega) &=&
 \frac{3}{4} \log \left(\cosh X(\omega)\right)-\frac{3 \beta  \arctan\left(\sinh X(\omega)\right)}{\bp^2},\nonumber \\
n_s(\omega)&=& 1-\frac{8 \bp^4 \cosh ^2X(\omega)}{\left(\bp^2 \sinh X(\omega)-4 \beta \right)^2}\nonumber \\
&& +\frac{8 \bp^2 \left(\bp^2 \cosh \left(2X(\omega)\right)-4 \beta  \sinh X(\omega)\right)}{3 \left(\bp^2 \sinh X(\omega)-4 \beta \right)^2}, \nonumber\\
 r(\omega)&=&\frac{64\bp^4\cosh^2 X(\omega)}{3 \left(\bp^2 \sinh X(\omega)-4 \beta \right)^2},\nonumber
\\
\hspace{-9cm}P_R(\omega)&=&\hspace{-0.4cm}\frac{\left(\beta -\frac{ \bp^2\sinh X(\omega)}{4} \right)^2 \left(\bp^2 \sinh X(\omega)-4 \beta \right)^2 \text{sech}^2X(\omega)}{128 \pi ^2 c \bp^8}. \nonumber
\eea}

Moreover, the  analytic expressions of $\omega_\pm$ (the two solutions of $\epsilon(\omega_{\rm end}) = 1$) are  
\bea \omega_\pm &=&\sqrt{\frac{3}{2}} \bp \left(\sinh^{-1}\left(\pm\sqrt{\frac{192 \beta ^2}{\bp^4}-4}-\frac{12 \beta }{\bp^2}\right) \right. \nonumber \\
&&\left.-\tanh ^{-1}\left(\frac{4 \beta }{\sqrt{16 \beta ^2+\bp^4}}\right)\right).\nonumber\eea


\begin{thebibliography}{00}
\bibitem{etaab}
$\eta_{ab}$ is the Minkowski metric, which we use to lower and raise the indices $a,b$.
\end{thebibliography}
\end{document}